\documentclass{article}
\usepackage{spconf,amsmath,graphicx}
\usepackage{psfrag}

  \usepackage[absolute]{textpos}
\newcommand{\copyrightstatement}{
    \begin{textblock}{15}(0.5,0.3)    
         \noindent
         \centering
         \textblockcolour{white}
         \footnotesize
         \copyright 2019 IEEE. Personal use of this material is permitted. Permission from IEEE must be obtained for all other uses, in any current or future media, including reprinting/republishing this material for advertising or promotional purposes, creating new collective works, for resale or redistribution to servers or lists, or reuse of any copyrighted component of this work in other works
    \end{textblock}
}

\title{Efficient Coding of $\mathbf{360^\circ}$ Videos Exploiting \\ Inactive Regions in Projection Formats}
\name{Christian Herglotz$^1$, Mohammadreza Jamali$^1$,  St\'ephane Coulombe$^1$, Carlos Vazquez$^1$, Ahmad Vakili$^2$ }
\address{$^1$\'Ecole de technologie sup\'erieure, Montr\'eal\\ $^2$ Summit Tech Multimedia, Montr\'eal
\vspace{-.5cm}}
\begin{document}
%

\copyrightstatement

\maketitle
\begin{abstract}
This paper presents an efficient method for encoding common projection formats in 360$^\circ$ video coding, in which we exploit inactive regions. These regions are ignored in the reconstruction of the equirectangular format or the viewport in virtual reality applications. As the content of these pixels is irrelevant, we neglect the corresponding pixel values in rate-distortion optimization, residual transformation, as well as in-loop filtering and achieve bitrate savings of up to $10\%$. 
\end{abstract}
\section{Introduction}
\label{sec:intro}
In recent years, virtual reality (VR) applications went through a remarkable evolution such that nowadays, many portable devices can process these applications in real time. A common example for a VR application is $360^\circ$ video playback. In this application, the user wears a headset with the portable device attached, where the display is used to create a simulated environment in which the user immerses. Turning his head in all directions, he can then see what happens around him in that environment. 

In this application, video data needs to be recorded and saved in a special projection format to cover all potential viewing angles \cite{Nielsen05}. This video data is commonly called $360^\circ$ video and overcomes the drawback of classic equilinear sequences, which usually cover less than $180^\circ$ of viewing angle horizontally and vertically, in a single point of view. 
To this end, a high number of different projection formats have been proposed and studied \cite{JVET_E1030, Yoon18, Kim18}. The target of a projection format is to optimize the user experience which is measured in terms of the visually perceived quality, independent from the viewing angle \cite{Xiu17}. 

In all these projection formats, the visual data is packed into a rectangular pixel array that can then be compressed with standard compression standards like H.264 \cite{ITU_H.264} or HEVC \cite{ITU_HEVC}. After decoding, the pixel data is used to undo the projection by calculating the underlying sphere or the viewport, which is then shown to the user. This method provides the advantage that most modern smartphones and portable devices are capable of decoding the visual data and performing subsequent reconstruction in real time  because dedicated video decoding hardware is available. 


Depending on the projection format, the projected visual data can have different properties. For example, the common equirectangular projection (ERP) distorts straight lines in a conventional equilinear  projection to curves \cite{Nielsen05}. If a projection is used that keeps straight lines like Cubemap (CMP), the video data is projected onto rectangular or triangular faces. These faces are packed into the rectangular pixel array such that either discontinuities appear at the face borders, or regions with inactive sample data occur that are not needed for the proper reconstruction of the video. In \cite{Jamali19a}, the effect of different projections on compression efficiency and computational complexity is studied.

This paper tackles the encoding method for projection formats containing inactive samples. Therefore, we modify the encoding process as follows. First, during the rate-distortion optimization (RDO), which determines the best prediction and coding parameters \cite{Sullivan98}, we propose to neglect the distortion imposed by pixels in inactive  regions. 
Second, we propose to set the inactive  residual coefficients to zero. 
Third, we neglect the inactive samples when collecting statistics for the best sample adaptive offset (SAO) post-processing filter \cite{Fu12}. 
The evaluation shows that without degrading the visual quality, average bitrate savings between $0.3\%$ and $6\%$ can be achieved in HEVC, depending on the projection format. 

This paper is organized as follows. First, Section \ref{sec:Irr} reviews projection formats and presents related work on $360^\circ$ video coding. Afterwards, we present the proposed coding method in detail in Section \ref{sec:algo}, show the evaluation setup in Section \ref{sec:eval}, and evaluate the performance in Section \ref{sec:RD}. Finally, Section \ref{sec:concl} concludes this paper.

\section{Inactive  Regions in $\mathbf{360^\circ}$ Videos}
\label{sec:Irr}
In the beginning of $360^\circ$ video coding, only few projection formats were considered. For example, in standardization, the first attempts to standardize $360^\circ$ video handling only considered ERP and CMP \cite{Skupin17}. Later on, with the establishment of the joint video exploration team (JVET), more projection formats were developed and studied to find the one best suited for efficient compression in the rate-distortion sense \cite{JVET_E1030}. Therefore, a dedicated software was written that is able to convert between all considered projection formats \cite{JVET_K1004} and evaluate their compression performance in a unified manner. 

\begin{figure}
\centering
\psfrag{O}[c][c]{\large{OHP}}
\psfrag{C}[c][c]{\large{COHP}}
\includegraphics[width=.4\textwidth]{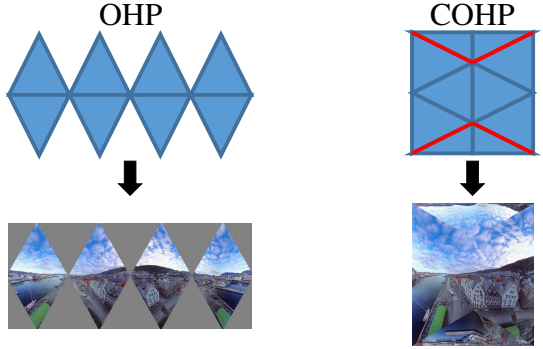}
\vspace{-.3cm}
\caption{Packing of triangular faces in octahedron projection. On top, the location of the faces is shown, the bottom shows an example of a projected frame of the AerialCity sequence.  }
\label{fig:exampleProj}
\vspace{-.3cm}
\end{figure}

For all the projection formats considered in this paper, the visual information is projected onto multiple faces. These faces are then packed into a rectangular frame. An example for two different packing methods 
is depicted in Fig.\ \ref{fig:exampleProj}. It shows octahedron projection (OHP), in which the input sphere is projected onto an octahedron consisting of eight triangles. 
On the left, a rectangular format is obtained by padding the spaces around the OHP faces with gray pixels. In compact octahedron (COHP) on the right, the triangles are packed more densely such that no gray area remains. However, on some borders between the triangles, the content is discontinuous as indicated by the red lines. In this case, $16$ samples are added between the faces to smooth the transition and hence reduce the coding bitrate \cite{JVET_K1004}. In the proposed algorithm, both kinds of inactive pixels are considered to improve the compression efficiency. 


The segmented sphere projection (SSP) segments the sphere into three tiles: north pole, equator and south pole \cite{Yoon18}. The boundaries of the three segments are $45^\circ$N and $45^\circ$S. The top and bottom segments (north and south poles) are mapped into two circles. The middle part (the equatorial segment) follows the same mapping as ERP and is projected onto a rectangle. The diameter of the circle is equal to the width of the equatorial segment because both polar segments and equatorial segment have a $90^\circ$ latitude span \cite{Yoon18}. SSP spends fewer pixels at the polar segments compared to ERP which helps improving the coding efficiency. However, the circular faces are padded with inactive samples. 

In icosahedron projection (ISP), there are $20$ triangular faces which are packed in a compact format (CISP). 
To compact the $20$ triangles in a rectangular frame, some triangles are split or flipped vertically or horizontally \cite{Kim18}. 

The rotated sphere projection (RSP) consists of two faces which are compacted in a rectangular frame in two rows. The top face is obtained from the middle part of ERP which includes the range of $270^\circ$ horizontally and $90^\circ$ vertically. The bottom face covers the same range. However, the basic sphere to obtain ERP is 
rotated $180^\circ$ along the Y-axis and $90^\circ$ along the X-axis.


Various works improve padding techniques for the inactive regions to increase the compression performance. He et al. propose a dedicated technique for discontinuous borders and report bitrate reductions between $0.3\%$ and $4.3\%$ \cite{He17}. Yoon et al. propose a technique for SSP \cite{Yoon18} with bitrate reductions of $0.3\%$.  Kim et al. propose a suitable technique when introducing the icosahedron projection \cite{Kim18}. 

Other work explicitly targeting discontinuous borders was done by Sauer et al. \cite{Sauer18}. They target the deblocking filter in CMP 
and modify the process when the block border is located on a discontinuity. They show that coding artifacts can be reduced significantly. Work targeting ERP projections was done by Budagavi et al. \cite{Budagavi15} performing region adaptive smoothing, in which up to $20\%$ of bitrate could be saved. In a similar direction, Li et al. proposed to use spherical-domain RDO to distribute the available bandwidth more evenly \cite{Li17} such that more than $10\%$ of bits could be saved at the same visual quality. To the best of our knowledge, the proposed method is the first that neglects the distortion of inactive samples during encoding.

\section{Proposed Algorithm}
\label{sec:algo}
To implement the proposed algorithm, we modify the HEVC reference software version HM-16.20 \cite{HM}. It is required to signal the encoder the locations of the inactive samples. As these locations do not change over time, it is sufficient to signal a binary mask at input resolution indicating whether a pixel is active ('1') or not ('0'). For each tested projection and resolution, such a mask is generated and signaled to the encoder. 
The following three subsections explain the modifications in distortion calculation, residual error transformation, and SAO handling, which can also be adapted for other coding standards. 

\subsection{Distortion Calculation}
In HM-16.20, rate-distortion optimization (RDO) is applied to find the best coding mode and coding parameters. Depending on this mode, three different kinds of distortion metrics are calculated: the sum of absolute differences (SAD), the sum of squared differences (SSD), and the sum of absolute Hadamard transformed differences (SATD). Assume that an original block $B$ and a distorted block $\tilde B$, defined for the set of pixel positions $\mathcal{P}$, are given. The distortion $D$ in terms of SSD is then calculated as 
\begin{equation}
D = \sum_{\mathbf{m} \in \mathcal{P}} \left(B(\mathbf{m}) - \tilde B(\mathbf{m})\right)^2, 
\end{equation}
with $\mathbf{m}$ a 2D pixel index.  

We assume that the block includes a set of inactive samples $\mathcal{I}$, which is a subset of $\mathcal{P}$. 
We now calculate the distortion as 
\begin{equation}
D_\mathrm{mod} = \sum_{\mathbf{m} \in \{\mathcal{P}\backslash \mathcal{I}\}} \left(B(\mathbf{m}) - \tilde B(\mathbf{m})\right)^2.  
\end{equation}
For SAD, the distortion calculation is modified accordingly. For SATD, before the Hadamard transform, the sample difference of inactive samples is set to zero. 


\subsection{Residual Coefficient Handling}
After prediction of a block, the residual error is determined and transformed with a discrete cosine transform (DCT) \cite{Sullivan12}, where 
the residual error is the difference between the original block and the predicted block. 
We propose a straightforward method to reduce the number of bits needed to code the residual coefficients in the DCT domain. 
The block of  spatial domain residual coefficients $R$ is calculated by 
\begin{equation}
R(\mathbf{m}) = B(\mathbf{m}) - \hat B(\mathbf{m}), \quad \forall \mathbf{m}\in \mathcal{P}, 
\end{equation}
where $B$ is the block of original pixel values and $\hat B$ the block of predicted pixel values. 

After this operation and before transformation, we set all values in the set of inactive samples $\mathcal{I}$ to zero as 
\begin{equation}
R(\mathbf{m})  = 0, \quad \forall \mathbf{m}\in \mathcal{I}.  
\end{equation}
This operation minimizes the power of the residual signal, which also minimizes the power of the signal in the transform domain such that fewer bits are needed to code the residual coefficients. 

\begin{figure}
\centering
\psfrag{E}[c][c]{\Huge{ERP}}
\psfrag{C}[c][c]{Conversion}
\psfrag{T}[c][c]{to Target}
\psfrag{M}[c][c]{Modified}
\psfrag{N}[c][c]{Encoder}
\psfrag{D}[c][c]{Decoder}
\psfrag{O}[c][c]{to ERP}
\psfrag{W}[c][c]{WS-PSNR}
\psfrag{R}[c][c]{Calculation}
\includegraphics[width=.38\textwidth]{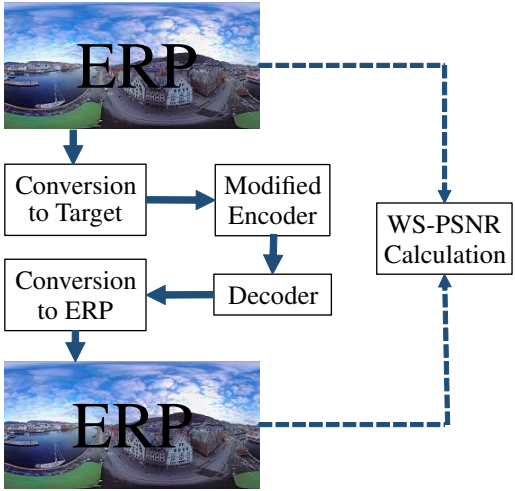}
\vspace{-.3cm}
\caption{Evaluation procedure for the proposed encoding method. All distortion calculations are performed in equirectangular domain. The target is one of the projection formats shown in Table\ \ref{tab:proj}.  }
\label{fig:evalSetup}
\end{figure}

\subsection{Sample Adaptive Offset}
In the HM-16.20 encoder, each coding tree unit (CTU) is tested for the sample adaptive offset filter (SAO) as introduced in \cite{Fu12}. Therefore, statistics are collected for all pixels by classifying them in the categories band offset (BO) and edge offset (EO), depending on their band and edge type. 

As explained in \cite{Fu12}, the band or the edge category 
of a pixel value 
is obtained as a function of the pixel value itself and adjacent pixel values. 
For edge offset, horizontal, vertical, or diagonal directions are considered. 
Then, for each category, a counter and the overall sum of the differences between the original and the reconstructed pixel values is defined. Depending on the categorization, the counter is incremented and the overall difference updated. 
The SAO parameters are then chosen to minimize the overall difference. 
In the proposed method, the categorization, the increment, and the addition to the overall difference are skipped if the pixel is located in an inactive  region. However, an inactive sample adjacent to an active sample can be used to determine the category of the active sample. 
\begin{table}[t]
\renewcommand{\arraystretch}{1.3}
\caption{Sequences used for algorithm evaluation. All sequences are coded in YUV420 format with a bit depth of $8$. The Balboa sequence has a frame rate of $60$ fps, the other sequences $30$ fps.  }
\label{tab:seqs}
\vspace{-.3cm}
\begin{center}
\begin{tabular}{c|c|l}
\hline
& Resolution & Sequences\\
\hline
$4$K & $3840\times 1920$ & AerialCity, PoleVault\\
$6$K & $6144\times 3072$  & Balboa, Landing2 \\
$8$K & $8192\times 4096$ & Gaslamp, Trolley \\
\hline
 \end{tabular}
\end{center}
\vspace{-.5cm}
\end{table}

\section{Evaluation Setup}
\label{sec:eval}
The framework depicted in Fig.\ \ref{fig:evalSetup} is used for performance evaluation.  
For conversion and distortion calculation, we use 360Lib \cite{360Lib} which introduces four new objective quality metrics in addition to the traditional PSNR: WS-PSNR, CPP-PSNR, S-PSNR-NN and S-PSNR-I \cite{JVET_E1030}. In this work, we use WS-PSNR to calculate the Bj{\o}ntegaard-Delta rate (BD-Rate) \cite{Bjonte01} which shows the average bitrate savings. 

WS-PSNR calculates the PSNR based on distortions for all pixels on the rectangular frame. However, unlike traditional PSNR, distortions are weighted based on the pixels’ positions. The weights are calculated based on the spherical areas which each pixel take on the sphere. It should be noted that the two frames which are used for WS-PSNR computations must have the same resolution and projection format.

We tested six different projection formats on six different test sequences. The test sequences are taken from the JVET common test conditions \cite{JVET_E1030, JVET_K1012} and are originally provided in the ERP projection format. The main properties are listed in Table\ \ref{tab:seqs}. The projection formats are taken from \cite{JVET_K1004} and the conversion is performed using the 360Lib-7.0 \cite{360Lib}. Depending on the resolution of the input sequence, different output resolutions are chosen as shown in Table  \ref{tab:proj}. 

\begin{table}[t]
\renewcommand{\arraystretch}{1.3}
\caption{Projection formats used for algorithm evaluation. $4$K, $6$K, and $8$K in the first line refer to the resolution of the input sequence (see Table\ \ref{tab:seqs}). Below, the listed resolutions represent the target resolution after projection, and the percentages show the fraction of inactive samples.   }
\label{tab:proj}
\vspace{-.6cm}
\begin{center}
\begin{tabular}{l|c|r||c|r}
\hline
 & \multicolumn{2}{c||}{$4$K } & \multicolumn{2}{c}{$6$K \& $8$K}\\
Proj. & Resolution & Inact.  & Resolution &  Inact.\\
\hline
CMP & $3840\times 2880$& $50.0\%$ & $4736\times 3552$ & $50.0\%$\\
OHP & $2880\times 1248$ & $49.7\%$ & $6176\times 2672 $ & $49.9\%$ \\
COHP & $2176\times 2552$ & $1.25\%$& $2672\times 3128 $ & $1.02\%$\\
 CISP & $1416\times 1816$& $8.71\%$ & $2496\times 3320 $ & $4.94\%$\\
 RSP & $2880\times 1920$& $5.33\%$ & $3552\times 2368 $ & $5.48\%$\\
SSP & $1008\times 6080$& $7.64\%$ & $1216\times 7328 $ & $7.56\%$ \\
\hline
 \end{tabular}
 \vspace{-.8cm}
\end{center}
\end{table}

For encoding the projected formats, we use HM-16.20 with the QPs $22$, $27$, $32$, and $37$ in the random access configuration \cite{Bossen13}. Because of a high simulation time, for all sequences, we code $33$ frames to cover two groups of pictures (GOPs) and two I-frames.

\section{Rate-Distortion Performance}
\label{sec:RD}
\begin{figure}
\centering
\psfrag{0000000000000}[cl][cl]{Proposed }%
\psfrag{00000001}[cl][cl]{Standard }%
\psfrag{002}[ct][ct]{ $0$}%
\psfrag{003}[ct][ct]{ $5$}%
\psfrag{004}[ct][ct]{ $10$}%
\psfrag{005}[rc][rc]{ $34$}%
\psfrag{006}[rc][rc]{ $36$}%
\psfrag{007}[rc][rc]{ $38$}%
\psfrag{008}[rc][rc]{ $40$}%
\psfrag{009}[c][c]{}
\psfrag{010}[c][c]{Bitrate [Mbps]}
\psfrag{011}[c][c]{WS-PSNR [dB]}
\includegraphics[width=.35\textwidth]{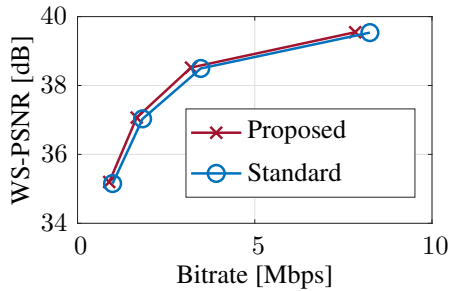}
\vspace{-.4cm}
\caption{Rate-distortion curves for the standard (blue) and the proposed (red) encoder. The sequence is AerialCity, the projection format is OHP, and the BD-Rate is $-9.77\%$. }
\label{fig:RD}
\vspace{-.3cm}
\end{figure}

Fig.\ \ref{fig:RD} shows the rate-distortion curves  for the AerialCity sequence in OHP projection. 
It can be seen that the curve corresponding to the proposed approach reaches the same WS-PSNR as the standard approach at a smaller bitrate. 

The results for all sequences and all projection formats are summarized in 
Table \ref{tab:results}. 
\begin{table}[t]
\renewcommand{\arraystretch}{1.3}
\caption{BD-Rate savings in terms of WS-PSNR [\%].  }
\vspace{-.3cm}
\label{tab:results}
\begin{center}
\begin{tabular}{l|c|c|c|c|c|c}
\hline
 & \multicolumn{6}{c}{Projection Format}\\
Sequence & CMP & OHP & COHP & CISP & RSP & SSP \\
\hline
AerialCity & $0.60$ & $ 9.77 $ & $ 0.85$ & $ 3.05$ & $0.63 $ & $0.94 $ \\
PoleVault & $ 0.15$ & $4.10 $ & $0.36 $ & $2.16 $ & $0.02 $ & $ 0.23$ \\
Balboa & $  0.44  $ & $8.16 $ & $1.55 $ & $2.76 $ & $ 0.41$ & $0.89 $ \\
Landing2 & $ 0.26 $ & $ 6.91$ & $1.65 $ & $3.22 $ & $0.48 $ & $1.24 $ \\
Gaslamp & $ 0.19  $ & $4.78 $ & $0.76 $ & $1.16 $ & $0.26 $ & $0.37 $ \\
Trolley & $ 0.15  $ & $2.39 $ & $0.36 $ & $0.78 $ & $0.09 $ & $ 0.21$ \\
\hline
Average &  $  0.30    $ & $ 6.02$ & $0.92 $ & $2.19 $ & $0.32 $ & $0.65 $ \\
\hline
 \end{tabular}
 \vspace{-.5cm}
\end{center}
\end{table}
We can see that on average, all projection formats achieve significant bitrate savings. Highest savings can be observed for the OHP projection (up to $10\%$). The reason is that in this projection format, the border between active and inactive samples crosses the coding block structure diagonally. As a consequence, the prediction needs to compromise between an accurate prediction of the edge and an accurate prediction of the active area, which in general leads to large residuals that are costly to code. The proposed method removes the need to reconstruct the edge such that bitrate can be saved. 

\begin{figure}[t]
\centering
\psfrag{S}[c][c]{a) Standard Encoder}
\psfrag{i}[c][c]{b) Proposed Encoder}
\includegraphics[width=.38\textwidth]{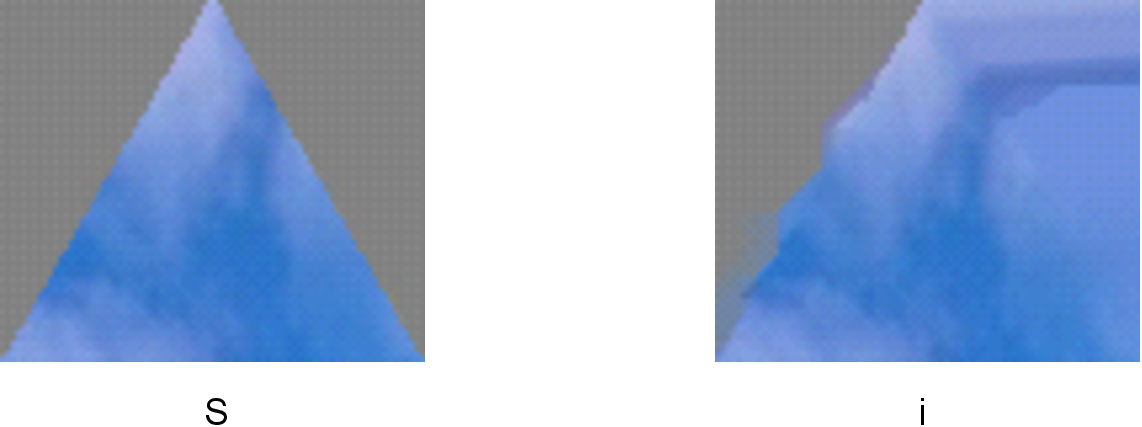}
\vspace{-.3cm}
\caption{Zoomed samples of the AerialCity sequence in OHP format after decoding for the standard (a) and the proposed encoder (b). The cutouts are taken from the first frame, top left face, coded with QP $22$. The face shows the sky.  }
\label{fig:comp}
\vspace{-.3cm}
\end{figure}

A visual comparison between the visual data supports this finding (Fig.\ \ref{fig:comp}). 
One can see that the original method reconstructs the gray area with a high quality. In the proposed method, the corresponding region on the right mainly shows predicted pixel values from an angular prediction mode. Hence, bits to code the residual are saved.

\section{Conclusions}
\label{sec:concl}
This paper presented an encoding method that disregards pixels in inactive  areas. The method is applicable to many $360^\circ$ projection formats and can be used for other formats in which inactive areas occur. Our evaluation indicates that the proposed method provides bitrate savings up to $10\%$. 

In future work, inactive regions can be exploited for encoder speedup. Studying the separate impact of distortion, residual, and SAO handling would also be interesting. Further projection formats like fisheye video can also be tested.

\section*{Acknowledgment}
This work was supported by Mitacs Canada and Summit Tech Multimedia (https://www.summit-tech.ca/). 

\bibliographystyle{IEEEbib}
\bibliography{IEEEabrv,literatureNeu}
\end{document}